\documentclass{aastex}

\shorttitle{Globular Cluster M92}
\shortauthors{Lee et al.}

\begin{document}

\title{Wide field CCD photometry of the globular cluster M92}

\author{Kang Hwan Lee}
\affil{Astronomy Program, SEES, Seoul National University, Seoul 151-742, Korea}
\email{khlee@astro.snu.ac.kr}

\author{Hyung Mok Lee}
\affil{Astronomy Program, SEES, Seoul National University, Seoul 151-742, Korea}
\email{hmlee@astro.snu.ac.kr}

\author{Gregory G. Fahlman}
\affil{Canada-France-Hawaii Telescope Cooperation, Hawaii 96743, USA}
\email{fahlman@cfht.hawaii.edu}

\and

\author{Myung Gyoon Lee}
\affil{Astronomy Program, SEES, Seoul National University, Seoul 151-742, Korea}
\email{mglee@astrog.snu.ac.kr}

\begin{abstract}
We present wide field CCD photometry of
a galactic globular cluster M92 obtained in the V and I bands with the 
CFH12K mosaic CCD at the Canada-France-Hawaii Telescope.
A well-defined color-magnitude diagram is derived down to 5 magnitudes fainter than
the cluster main sequence turn-off.
After removing the background contribution, we obtain luminosity and mass functions, surface density profiles, and the surface number density maps
of the stars belonging to the cluster.
The surface density profile of all stars shows that the cluster's halo extends 
at least out to
$\sim30'$ from the cluster center in agreement with previous study,
but the profile of faint stars at the very outer region of the cluster 
shows a different gradient
compared with that of bright stars.
For a mass function of the form $\Phi(M) \propto M^{-(1+x)}$, we find that 
the inner region ($5' < r < 9'$) of the cluster has $x\simeq1.2\pm0.2$, 
whereas the outer region ($9' < r < 15'$)
has $x\simeq1.8\pm0.3$, clearly indicating a mass segregation of the cluster.
An estimate of the photometric mass of the cluster implies that 
the remnant populations (white dwarfs and neutron stars) contribute at least
25\% of the total cluster mass.
The surface density map of M92 shows some evidence that the tidal tail of M92 may
be oriented perpendicular to the direction toward the
Galactic center.

\end{abstract}


\section{INTRODUCTION}

Globular clusters are good laboratories for testing theories of stellar dynamics 
and galactic structure.
All globular clusters ultimately decay, through a combination of internal
(two body relaxation, and stellar evolution) and external 
(gravitational shocks due to the
Galactic bulge and disk, and dynamical friction) processes (Aguilar, Hut,
\& Ostriker 1988).
Lee \& Goodman (1995) and Gnedin \& Ostriker (1997) predicted 
that possibly as many as half of the 
present-day Galactic globular clusters would be disrupted in another Hubble time.
Many previous studies have shown that tidal shocks play a very important role
in cluster evolution by accelerating the destruction of the cluster 
(e.g. Ostriker, Spitzer, \& Chevalier 1972;
Gnedin \& Ostriker 1997;
Gnedin, Lee, \& Ostriker 1999).
Lee \& Ostriker (1987) predicted that, for a given tidal field, the outer structure
of a globular cluster will be more populated than that of a King model, since the
density at the tidal radius would not be exactly zero.
In addition, the cluster is expected to
have tidal tails because the equipotential surface is not spherical.
The presence of tidal tails from globular clusters has 
begun to be explored only recently because of difficulties in obtaining accurate
wide field photometry.
The development of computing power and wide field digital detectors now
allows us to study these very interesting parts of globular clusters in detail.
Indications of tidal tails around globular clusters were found
in several previous studies of globular cluster radial density profiles.
Grillmair et al. (1995) examined the outer structure of 12 Galactic globular 
clusters, and found that 10 of their sample clusters showed extra-tidal
wings in their surface density profiles.
Later, Lehmann \& Scholz (1997) reported the discovery of tidal tails in 5 out of 7
globular clusters.
Evidence of tidal extensions are even found in four globular clusters in M31 
(Grillmair 1996; Holland, Fahlman, \& Richer 1997).
Leon, Meylan, \& Combes (2000) investigated 20 Galactic globular clusters with Schmidt
plates and films, and found that all of the clusters that do not
suffer from strong observational biases, show tidal tails.
On the theoretical side, Combes, Meylan, \& Leon (1999) used N-body simulations to show 
that two giant tidal tails are expected to exist along the galactic orbit of a 
globular cluster.
Yim \& Lee (2002) also examined the general evolution of the clusters using
N-body simulation and showed that the cluster density profiles
appear to become somewhat shallower just outside the tidal boundary,
and that the directions of the tidal tails depend on the location in the
Galaxy as well as the cluster orbit.

Recently, Odenkirchen et al. (2001) reported the discovery of two well defined
tidal tails emerging from the sparse, remote globular cluster Palomar 5.
They used the CCD images from Sloan Digital Sky Survey (SDSS).
Except for this study (and those of the clusters in M31), 
all the previous observational works on globular
cluster tidal tails were based on data from photographic plates.
However, such data generally are not deep enough to examine the low mass stars
which dominate the outer parts of globular clusters, 
and the photographic photometry
tends to show greater scatter than that based on CCD data.

In the specific case of M92, Testa et al. (2000) investigated the dynamical 
structure using plates from the Digitized Second Palomar Sky Survey (DPOSS).
Although they found some evidence for an elongation in the extra-tidal
extension from their surface density map, the significance of their 
result was low. In this paper, we report the results of a similar analysis 
using deep CCD images. We surveyed
 a $2^\circ \times 2^\circ$ region
centered on M92 with CFH12K in order to study the dynamical structure
and tidal tails of the cluster.
We also performed a very detailed analysis of the main sequence mass 
function, which was impossible to do in the study using photographic plates.

M92 (NGC 6341)
is a very metal poor ([Fe/H] = $-2.24$)globular cluster located
at a distance from the galactic center of
$R_{GC} = 9.1$ kpc, and is 
$Z_{GC} = 4.3$ kpc
above the galactic plane (Harris 1996).

This paper is organized as follows.
Sect. 2 presents the data and the reduction procedures.
In Sect. 3, we show CMDs and discuss the blue straggler stars.
The radial profiles of stars in different magnitude ranges
are obtained in Sect. 4.
The resulting LFs and MFs are presented and discussed in Sect. 5,
and the photometric mass is estimated in Sect. 6.
In Sect. 7, the surface density maps are discussed.
Final results are summarized in Sect. 8.

\section{OBSERVATION AND DATA REDUCTION}

The observations were made with the 3.6 m Canada-France-Hawaii Telescope (CFHT)
during 2000 June 27-29, using CFH12K, a $6 \times 2$ 
mosaic of $2048\times 4096$ CCDs.
The instrument has an angular scale 
of $0.\arcsec 206$/pixel at the f/4 prime focus of the CFHT,
giving a field of view of $42\arcmin \times 28\arcmin$.
The CFH12K camera is well suited for this study as it covers a large area on
the sky with well sampled images.
A total of 10 fields through V and I filters were observed covering a field of
$120\arcmin \times 120\arcmin$, significantly larger than
the tidal radius of M92, $15.\arcmin2$, estimated by 
Trager et al. (1995).
The positions of the observed fields relative to the center 
of M92 are shown in Fig.~1, and pertinent observational
information is given in Table~1. All of the science images were 
obtained under good seeing
conditions (FWHM of $\sim 0.\arcsec8$). 
We also obtained frames of fields containing Landolt (1992) standard stars
for photometric calibration, and several twilight (flat-field), 
bias and dark frames were also taken to permit removal of
the instrument signature from the science data.

\placefigure{fig1}

\placetable{tab1}

All pre-processing, bias and dark subtraction and flat fielding, was 
performed using the FITS Large Images Processing software (FLIPS),
a very efficient package developed 
by Jean-Charles Cuillandre for the reduction of CCD mosaic images
(see Kalirai et al. 2001 for more information).
Instrumental magnitudes of the point sources in the images 
were derived using the programs DAOPHOT II/ALLSTAR (Stetson 1994).
The point-spread function (PSF) is different for each of 
the 12 CCDs on the mosaic, so the analysis is done separately for each chip.
In those cases where we have 3 images in the same band 
(F3, F4, F5, F6 and F7 in V),
we averaged them 
using the ALIGN and IMCOMBRED commands within FLIPS.
For the other fields, we analyzed each image separately 
and then averaged the photometry to get
mean instrumental magnitudes.
We used the stellarity index of SExtractor (Bertin \& Arnouts 1996) 
for separating stars from background galaxies and, as an example, 
Fig.~2 shows this parameter for sources detected in F3.
Those objects with a stellarity index greater than 0.8 were 
considered to be stars.
This process has excluded most of the background galaxies.
After further cuts, based on
 the two DAOPHOT parameters, the magnitude error ($< 0.2$) and the 
PSF fitting parameter, $\chi^2$ ($< 2$), were made to eliminate spurious 
measurements, the data sets in each
filter were matched and the positions, instrumental magnitudes,
and colors were obtained for all the stars.

\placefigure{fig2}

The instrumental magnitudes were then transformed to the standard
Johnson-Cousins
photometric system by using the photometric standard stars in
SA113
(Landolt 1992).
The standard values for the color term and the atmospheric 
extinction coefficient given on the CFHT home page were used to
to find the zero points of each chip separately. The
instrumental zero points for the chips were almost identical
(within a range of 0.1 mag)
because FLIPS normalizes the background sky value to the chip with
the highest sky value (lowest gain).
Fig.~3 shows the differences between our photometry of the Landolt (1992)
standard stars and the standard values.

\placefigure{fig3}

\section{The COLOR-MAGNUTUDE DIAGRAM}

The derived CMD for the central region of the cluster is shown in Fig.~4.
and is based on the data obtained by merging the long and short exposure data.
The short exposure data were used to cover the center ($r < 2'$)
of the cluster where bright stars were totally saturated in long exposure data,
and to show the evolved sequences of the cluster.
The inner core of the cluster ($r < 1'$) is too crowded to resolve stars even in 
short exposure data.
The solid line is a fiducial line from the previous 
study by Harris et al. (1997), showing a good agreement between the two.
Our primary concern is the properties of the cluster main sequence stars, 
so we will deal with
mainly this part of the CMD in the following.

The CMD of the distant fields was used for 
first-step of removing background/foreground field contamination.
For comparison we show CMDs of inner ($2' < r < 15'$) and
outer ($r > 45'$) regions separately in Fig.~5.
We selected stars within the illustrated color band ($\Delta (V-I) = 0.5$) centered on the
fiducial magnitude as candidate main sequence stars.
Our selection criteria is wide enough to include all
members of cluster main sequence but it will also include 
some contribution from field stars that can only be removed statistically.

Twelve stars in the box of Fig.~4 are thought to be blue straggler (BS) stars.
BS stars are known to be more centrally concentrated than others in globular
clusters (Bailyn 1995). In Fig.~6, we compared the radial distribution of
BS stars with that of subgiant branch (SGB) star.
Though the sample of BS stars is very small, we find no evidence that the
BS stars are more concentrated than SGB stars.
But further discussions on the BS stars are beyond the scope of this study.

\placefigure{fig4}

\placefigure{fig5}

\placefigure{fig6}

\section{THE RADIAL DENSITY PROFILE}

\subsection{Incompleteness Correction}

The stars within the specified range in the CM diagram are likely to be members
of the cluster.
However, because of crowding effect, we cannot recover all the members of cluster.
In order to study the spatial distribution of stars, we need to correct for
the crowding. 
This is called incompleteness correction.
Usually, the completeness depends on the level of crowding which depends on
location with respect to the cluster center.
An increasing contribution of the sky background also reduces the
detection probability for faint stars.
To correct for incompleteness, we ran artificial star tests
on the V frames, using DAOPHOT/ADDSTAR.
We chose 22 chips in four directions from the cluster center.
The tests were applied on each chip separately.
First, we added artificial stars in each 0.5 mag bin randomly
on original images.
The number of added stars was designed not
 to exceed 10\% of the total number of stars
that is actually present in that bin so as not to enhance original crowding.
The new frames obtained in this way were then reduced in an
identical manner used on the original frame.
In Fig.~7, we showed the plots of the differences between input and output
magnitudes for recovered artificial stars selected by the same criteria
which were applied to original image reduction.
The stars, which have magnitude differences smaller than
0.3 mag are considered as recovered stars.
We repeated these procedures 10 times on each chip to obtain meaningful
statistics in each magnitude bin.
Finally, the incompleteness correction factor $f$ is obtained by
$f = n_{rec}/n_{add}$,
where $n_{add}$ is the number of added stars and $n_{rec}$ is the 
number of recovered stars.
The resulting incompleteness correction factors applied to correct
luminosity functions are listed in Table~2.

\placefigure{fig7}

\placetable{tab2}

\subsection{Field Subtraction}

Although we have used CMD selection to reduce background/foreground field
contribution, there are still many field stars within the defined main
sequence band on the CMD (Fig.~5).
To build accurate radial density profiles and luminosity functions
 of the cluster, this contamination needs to be statistically removed.
Since the length of tidal tails is an order of 5 tidal radii,
or greater (Combes et al. 1999), which is larger than the region surveyed here,
a determination of the field density using
all the outermost regions could lead to an overestimate.
Therefore, in order to estimate the backgroun level as a first step,
we selected those regions showing no overdensities in the outermost parts
in each magnitude bins after constructing the surface densigy maps
(see section 6 for details).
The stellar densities in the selected regions were then averaged with weighting factors
proportional to the area of the selected regions.
The determined background densities are $0.18\pm0.03$arcmin$^{-2}$ 
for $18.5 < V < 20.5$, $0.40\pm0.04$arcmin$^{-2}$ for $22 < V < 23.5$,
and $0.85\pm0.05$arcmin$^{-2}$ for $18.5 < V < 23.5$.

\subsection{Radial Density Profile} 

The cluster's radial density profile was obtained by counting 
all the main sequence stars in annuli with widths of $2\sim3$ arcminutes.
These numbers were then corrected for field contamination and
incompleteness as
described in previous sections.
The final surface density of $i$'the annulus was then derived, using:
\[
SDP_i = -2.5 \times \log(N_{i,i+1} /A_{i,i+1}) + 2.5 \times \log f + C,
\]
where $N_{i,i+1}$ indicates the background subtracted 
number of stars in the annulus between
$r_i$ and $r_{i+1}$,
$A_{i,i+1}$ the area of the annulus and
$f$ the incompleteness correction factor.
The constant, C, was chosen to match the profile with that of 
Trager et al. (1995) in the overlapping region. 
The effective radius of each annulus is given by:
\[
r_i ^{eff} = \sqrt{{1 \over 2}  \times (r_i ^2 + r_{i+1} ^2 )}.
\]

The final radial density profile of M 92 using the entire main sequence
sample of stars ($18.5 < V < 23.5$) is shown in Fig.~8.
In the central region ($r < 2'$), the short exposure data were used with
smaller radial steps (10 arcseconds).
However, the points for the inner most region ($r < 1'$),
where the stars are not resolved, were obtained not by counting stars
but by surface photometry of the short exposure data. 
The flux measurements from surface photomtery were shifted to match with
star counts.
These points are shifted to match with the results from outer region.
The measured surface density profile is listed in Table~3.
In Fig.~8 we see that our result agrees very well with that 
of Trager et al. (1995). The best fitting single-mass isotropic 
King model profile is shown on Fig.~8 as a solid line 
(W$_0$ = 8, c = 1.85).
The tidal radius of the best-fit King model is $r_t = 840\arcsec$,
which lies between the $r_t = 740\arcsec$ by Testa et al. (2000),
and $r_t = 912\arcsec$ of Trager et al. (1995).
The data exhibits a noticeable deviation from the
King model in the outer region.
This is a sign of the presence of extra-tidal material.

\placefigure{fig8}

\placetable{tab3}

Grillmair et al. (1995) examined 
the density profiles of 12 Galactic globular clusters
 and found that most of them had extra-tidal stars. They described
the radial distribution of extra-tidal stars by a power law
$\Sigma (r) \propto r^{\gamma}$
with $-5 < \gamma < -1.6$. Using numerical simulations, 
Johnston, Sigurdsson, \& Hernquist (1999) showed  
that the extra-tidal population
is expected to have $\Sigma (r) \propto r^{-1}$,
and that there should be a break in the slope of
the surface density profile  
due to stars being stripped from the cluster.
On the other hand, Combes et al. (1999) deduced 
a slope around $\gamma = -4$ from their N-body simulations.
They explain the shallower slope in the observations as the contamination
of background-foreground populations.
Combes et al. (1999) also predicted that the escaped stars would not 
be distributed homogeneously within the tails, but form clumps that
mark the occurrence of the strongest gravitational shocks.
Leon et al. (2000) investigated tidal tails of 20 Galactic globular clusters
and computed the slopes of surface density of extra-tidal stars for their
sample. The values are in the range of previous studies 
$-5.03 < \gamma < -0.35$, but not all clusters
showed a break in the slope.

Testa et al. (2000) obtained the surface density profile of M 92 using
plates from the Digitized Second Palomar Sky Survey.
They fitted the extra-tidal profile of the cluster to a power law and found
$\gamma = -0.85\pm0.08$.
From our surface density profile, shown in Fig.~9,
we derive $\gamma = -0.82\pm0.10$ using only the
bright stars ($18.5 < V < 20.5$), which we take to correspond with 
the stars in the study by Testa et al. (2000).
Using the faint stars ($22 < V < 23.5$), we find a 
steeper slope of $\gamma = -1.31\pm0.09$,
which is significantly different from that of the bright stars.
Using the entire range of calibrated stars
($18.5 < V < 23.5$), we find $\gamma = -1.27\pm0.08$.
We see a clear clump in the extra-tidal profile 
(around $r = 2100\arcsec$),
a feature similar to those predicted by Combes et al. (1999).

It is noteworthy that the slopes of the density profiles 
for stars in different magnitude bins,
have different values in the outer part of the cluster.
Since the tidal tails of globular clusters are preferentially formed by the lowest
mass stars (Combes et al. 1999), it is difficult to study tidal tails of
globular clusters using only bright stars.
This implies that the study of globular clusters tidal tail should be performed
using stars with the lowest possible mass that is detectable.
It is possible that the difference in slopes of the density profiles in different magnitude bins is a real feature of the cluster.
Deep CCD studies of other nearby clusters are needed
to verify whether the mass dependence of the density profile slope
is a common phenomenon.

\placefigure{fig9}

\section{THE LUMINOSITY AND MASS FUNCTION}

A luminosity function (LF) was constructed by using the
main sequence stars within the cluster's tidal radius.
To investigate any spatial difference, the LFs for an
inner region ($5' < r <9'$) 
and an outer region ($9' < r < 15'$) was constructed separately.
The region of $r < 5'$ was excluded here to avoid the
uncertainty due to central crowding. As noted below, HST results are
available for the region inside $r < 5'$. 
To build correct LFs, we applied background subtraction
and incompleteness corrections as described in the previous sections.
The corrected number of stars in each magnitude bin 
and the incompleteness factors
are given in Table~2, and the results are plotted in Fig.~10. 
The histogram in Fig.~10 is the LF for all stars within $5' < r < 15'$.
We find that the outer LF is steeper than the inner LF, a clear sign
of mass segregation in the cluster.
The solid line overlaid on the inner LF is 
the LF of M92 for $r \sim 5'$    
derived from HST WFPC2 data by Piotto, Cool, \& King (1997).
Our result agrees very well with HST result.

The LFs were converted into mass functions (MFs) using the mass-luminosity
relation of Baraffe et al. (1997) and a distance modulus of
$(m-M)_V = 14.82$ (Gratton et al. 1997).
The calculated MFs are shown in Fig.~11.
Representing the MF by a power law,
\[
\Phi (M) dM \propto M^{-(1+x)} dM,
\]
we derive slopes of $x = 1.2 \pm 0.3$ and $x = 1.8 \pm 0.5$
for the mass 
range $0.35M_\odot < M < 0.8M_\odot$ in the inner($5' < r <9'$)
and outer ($9' < r < 15'$) regions,  respectively.
The slope for the entire region between $5' < r < 15'$ is
 $x = 1.3 \pm 0.2$. 
For the mass range $0.26M_\odot < M < 0.73M_\odot$, 
Andreuzzi et al. (2000) derived a 
slope of $x = 0.71$ between $2' < r < 5'$ from HST data.
From these results, we conclude that the                
slope of the MF increases with distance from the cluster center,
as expected from mass segregation in clusters.

McClure et al. (1986) suggested that the slope of MF is related to
the metallicity of cluster.
Djorgovski, Piotto, \& Capaccioli (1993) demonstrated that the MF slopes are determined
not only by the metallicity but also by the location in the Galaxy.
The strongest dependence is on the Galactocentric distance ($R_{GC}$): 
clusters closer to the Galactic center have flatter MFs.
At a given Galactocentric distance, clusters with a smaller distance
from the Galactic plane  ($Z_{GC}$) have flatter MFs, and at a given position,
clusters with lower metallicity have steeper MFs.
Capaccioli, Piotto, \& Stiavelli (1993) interpreted the dependence on position as
the effect of tidal shocks.
Disk and bulge shocking act preferentially on stars located in the 
outer region of the cluster and lead to a loss of low mass stars.
As a result the clusters near the Galactic plane,
tend to have flatter MFs.
Consequently, the relatively steep MF slope of M92 could be explained by
its low metallicity ([Fe/H] = $-2.24$) and large $R_{GC}$ (9.1 kpc) and
$Z_{GC}$ (4.3 kpc). 

Piotto et al. (1997) confirmed the relation between 
Galactocentric positions and MF slopes of
globular clusters by comparing LFs of four metal poor globular clusters.
They showed that three of four clusters (M15, M30, and M92) have nearly
identical LFs, whereas NGC 6397, with the smallest $R_{GC}$ and $Z_{GC}$, has 
a distinctly different LF, especially the fainter part.
They suggest that the three globular clusters, which have similar LFs, were
formed with similar MFs that have changed little (or changed in
similar ways), while NGC 6397 was more vulnerable to tidal shocks.
Lee, Fahlman, \& Richer (1991) showed that 
the slope of the cluster mass function can change rapidly only during the
final stage of dynamical evolution and that 
one signature of highly evolved clusters is a significant flattening
of the mass function (see also Lee \& Goodman 1995, and
Takahashi \& Lee 2000).
The relatively steep MF slope of M92, especially in the outer
part of the cluster, suggests that this cluster, although it has 
experienced substantial mass segregation, has been little affected 
by tidal shocks.

\placefigure{fig10}

\placefigure{fig11}

\section{MASS ESTIMATION}

Since we have found nearly all the stars in the cluster
outside of the central region with masses greater than 
$0.35M_\odot$, we can
estimate the total luminous mass of the cluster by 
combining our data with that in the HST archive for the cluster center.
HST archival data covering $r < 5'$ was obtained and analyzed  
using HSTphot (Dolphin, 2000). We counted stars within annuli of width
$14\arcsec$.  Since the HST archive data do not cover a full 
360 degree azimuthal annulus, we complete the counts by assuming that
the cluster has circular symmetry.
All the star counts are corrected for completeness and converted to masses
using the mass-luminosity relation of Baraffe et al. (1997).
The limiting magnitude, defined to be the value where 
the completeness fraction drops below 50\%, and
the calculated mass above the limiting magnitude 
is listed in Table~4 for each annulus. 
Summing these estimates, we
find that the measured luminous mass is $\sim 7.4\times 10^4 M_\odot$.
This value is clearly only a lower limit to the luminous stellar mass 
of the cluster. The total mass of the fainter, uncounted stars must be 
estimated in some way.

We can estimate the total mass of the low mass stars by extrapolating
the observed mass functions to the hydrogen burning limit, here taken
to be $0.08 M_\odot$. Explicitly, we assume that the observed MF slope 
applies right to this limit. 
This may not be an unreasonable assumption because 
recently, Richer et al. (2002) found that the mass function of the nearby                
globular cluster M4, which rises slowly toward their lowest observed mass of
0.09$M_\odot$, shows no convincing evidence of a turnover.
In any case, the lowest mass stars ($< 0.09 M_\odot$) contribute only 
small part of the cluster's total mass (below 5\%) even at the steepest slope
($x = 1.8$).
The extrapolated mass (that below the limiting magnitude) in each annulus is
 shown in Table~4.
The last column in Table~4 lists the slope of the 
mass function that was used to calculate the extrapolated mass. 
These values are based on the results of
Andreuzzi et al. (2000) for the region $r < 5'$ and present study for the
region $5' < r < 15'$.
As expected, the correction for uncounted stars is substantial 
within the inner arcminute of the cluster. 
The estimated total mass is $\sim 1.6\times 10^5 M_\odot$.
This value is arguably an upper limit to the total main-sequence mass for two reasons.
First, the mass function may not be constant to the limit of $0.08 M_\odot$, 
and second the MF slope in the core of the cluster ($r < 1'$) is expected to
be smaller than the value we used, which was derived from a somewhat more
distant region. The luminous mass we determine is smaller 
than the estimated dynamical mass of
$\sim 2.1\times 10^5 M_\odot$ by Pryor \& Meylan 1993.
The mass deficiency is probably due entirely to a remnant 
population in the cluster - white dwarfs and neutron stars.
Our derivation of the luminous mass gives an uncertain result for the reasons
discussed above and also because it depends on stellar models. 
The current theoretical models for low temperature
stars of low metallicity do not fit the lower main-sequence very well, leading to another
source of systematic uncertainty.
With these caveats, we cautiously predict that the remnant population
contributes at least 25\% to the total mass of M92.  

\section{SURFACE DENSITY MAP}

In order to characterize the distribution of the extra-tidal
stars, we constructed surface number density maps as follows.
The number counts of main-sequence stars in the
surveyed region were binned on a grid of $0.'5 \times 0.'5$.
We then convolved the map with a Gaussian kernel of width $2.'5$.
The resulting smoothed surface density maps are shown in Fig.~12.
The two maps are constructed using:
(a) bright stars (about 2 magnitudes below
the main-sequence turn-off), 
and (b) the entire set of stars, which is dominated by faint objects.
We overlaid contour levels and marked the tidal radius with
a thick circle.
The long arrow indicates the direction to the galactic center 
and the short arrow shows the proper motion 
($\mu_\alpha \cos\delta = -3.30\pm0.55$ mas yr$^{-1}$,
$\mu_\delta = -0.33\pm0.70$ mas yr$^{-1}$) compiled by Dinescu,
Girard, \& van Altena (1999).

From an inspection of the two maps, we see that the
shape of the tidal halo is different between the bright and faint stars.
In particular, a clump to the SE of the cluster appears in the bright star 
density map (it also appears in the Fig.~6 of Testa et al. 2000) is not seen in 
the map for faint stars.
It is difficult to see an elongation of the tidal halo with the 
bright stars only
because their number counts are typically too small.
Deep CCD photometry is needed to 
define the shape of the tidal halo, especially for the halo clusters
that is expected to have a weak extra-tidal extension.

From Fig.~12, we see a marginal elongation
orthogonal to the direction to the Galactic center (NE to SW).
Several models (Murali \& Dubinski 1999; Combes et al. 1999; Yim \& Lee 2002) predict
that there may exist two giant tidal tails around the globular cluster.
Combes et al. (1999) showed that the tidal tail follows the cluster orbit,
but Yim \& Lee (2002) showed that the directions of the tidal tails
depend on the location in the Galaxy as well as the cluster orbit.
Tidal tails have been detected around many globular clusters 
(Grillmair et al. 1995; Leon et al. 2000; Odenkirchen et al. 2001) and
some dwarf spheroidal galaxies of the Local Group (Mateo, Olszewski, \& Morrison 1998).
Possible explanations of the weak tidal halo of M92 are that the
tail may be compressed along the line of sight or the evaporating stars 
might not yet have been formed a tidal stream.
According to the numerical simulation by Yim \& Lee (2002), tidal tail is
longest near the apogee and shortest near the perigee. 
Therefor M92 might be near perigee.

\placefigure{fig12}

\section{SUMMARY}

We have carried out a wide field CCD survey of M92 and 
investigated the dynamical structure and stellar population 
of the cluster M92.
The main results of our study are:

(i) We find some candidates of BS stars in the CMD. 
We can not find any evidence that the BS stars are more centrally concentrated
than SGB stars.

(ii) The radial density profile of M92 reveals the presence of extra
tidal material.
From the King model fitting method, we derived a tidal radius for
M92 of $r_t = 840\arcsec$.
We find that the extra-tidal density profile is shallower for the bright
star when compared to the fainter stars.

(iii) We show that the
slope of cluster MF increases toward the outer part of the cluster, as expected from
mass segregation.
The relatively steep MF slope of M92, especially in the outer part of the
cluster, suggests that this cluster has not been strongly affected by
tidal shocks.
 
(iv) From an estimation of the photometric mass of M92, we conclude
that the maximum value of the cluster's MS mass is $\sim 1.6\times 10^5 M_\odot$.
At least 25\% of the total cluster mass may  be contributed by the
remnant population.

(v) From the surface density map, we see a slight elongation of
the cluster halo perpendicular to the
direction to the Galactic center.
Possible explanations of the weak tidal halo of M92 are
that the tidal tail may be compressed along the
line of sight,
or the evaporating stars may not yet have been formed a tidal stream.

Finally, we note the need for deeper CCD photometry to investigate
extra-tidal stars, especially for clusters, like M92, which have weak 
tidal extensions.



\clearpage

\figcaption[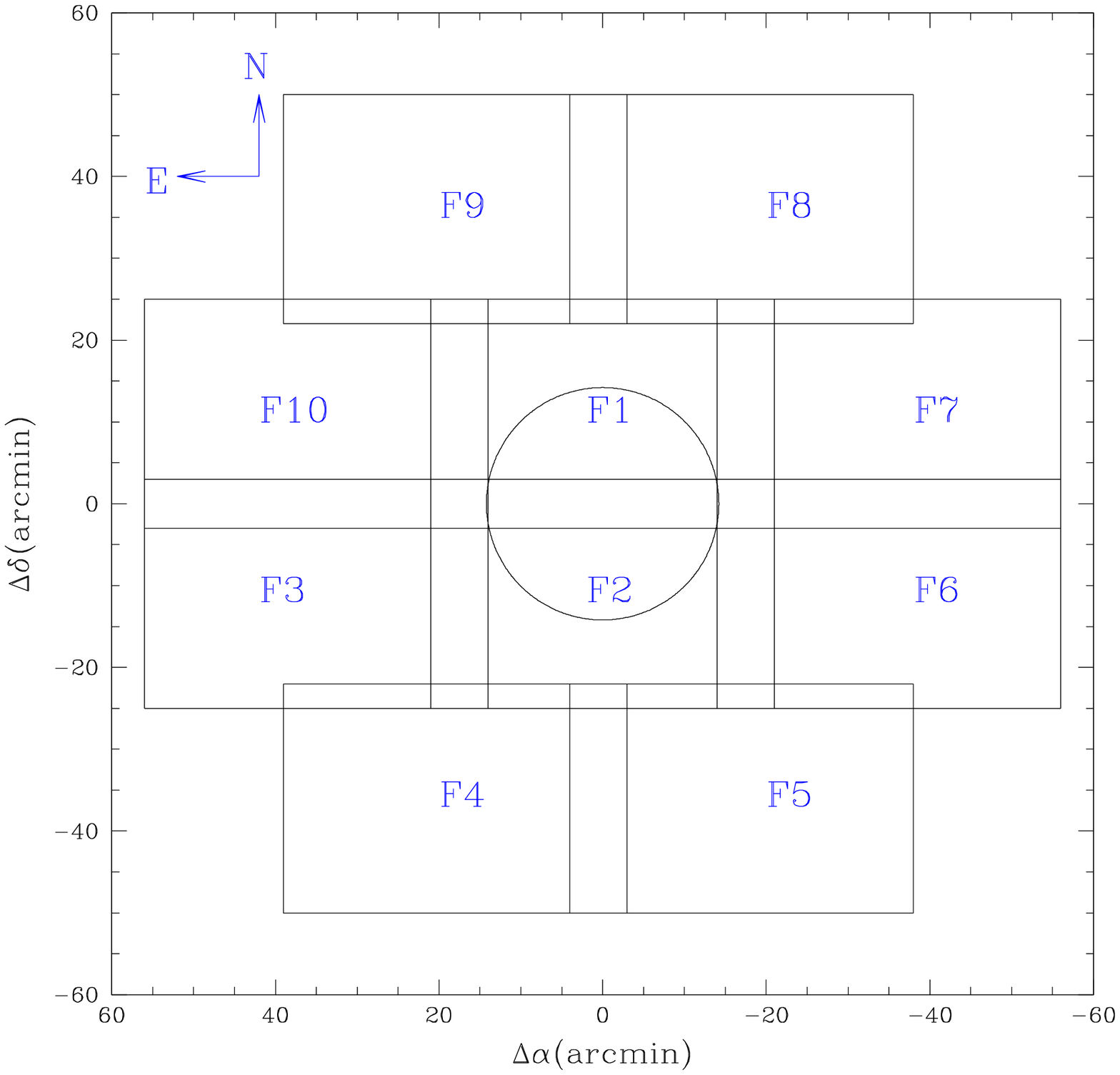]{Location of 10 observed fields, with the origin set on the
cluster center. 
The circle indicates derived tidal radius $r_t = 14'$.
Each frame has a field of view of $42' \times 28'$.
\label{fig1}}

\figcaption[Lee.fig02.ps]{Stellarity index of objects in F3 region.
\label{fig2}}

\figcaption[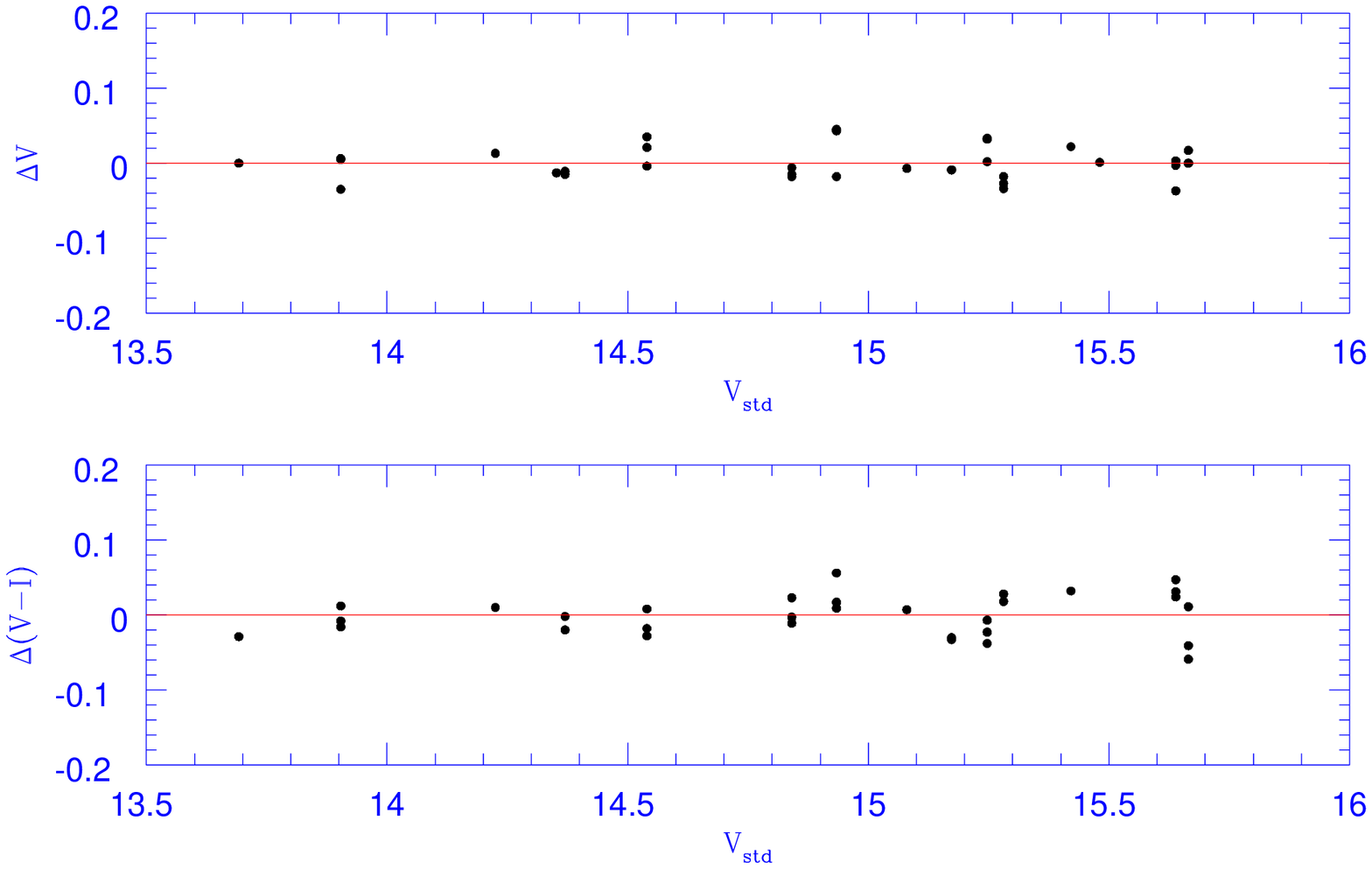]{Differences between our photometry of the Landolt (1992)
standards and their published values. The differences $\Delta V$ and $\Delta (V-I)$
are (Landolt's - ours).
\label{fig3}}

\figcaption[Lee.fig04.ps]{CMD for the central region of the cluster ($1' < r < 8'$).
This is the result of merging the long and short exposure data.
The solid line is a fiducial line from Harris et al. (1997).
\label{fig4}}

\figcaption[Lee.fig05.ps]{CMDs devided by inner ($2' < r < 15'$) and
outer ($r > 45'$) region of the cluster.
The overlayed lines indicate criteria for selecting cluster MS stars
to reduce field star contribution.
\label{fig5}}

\figcaption[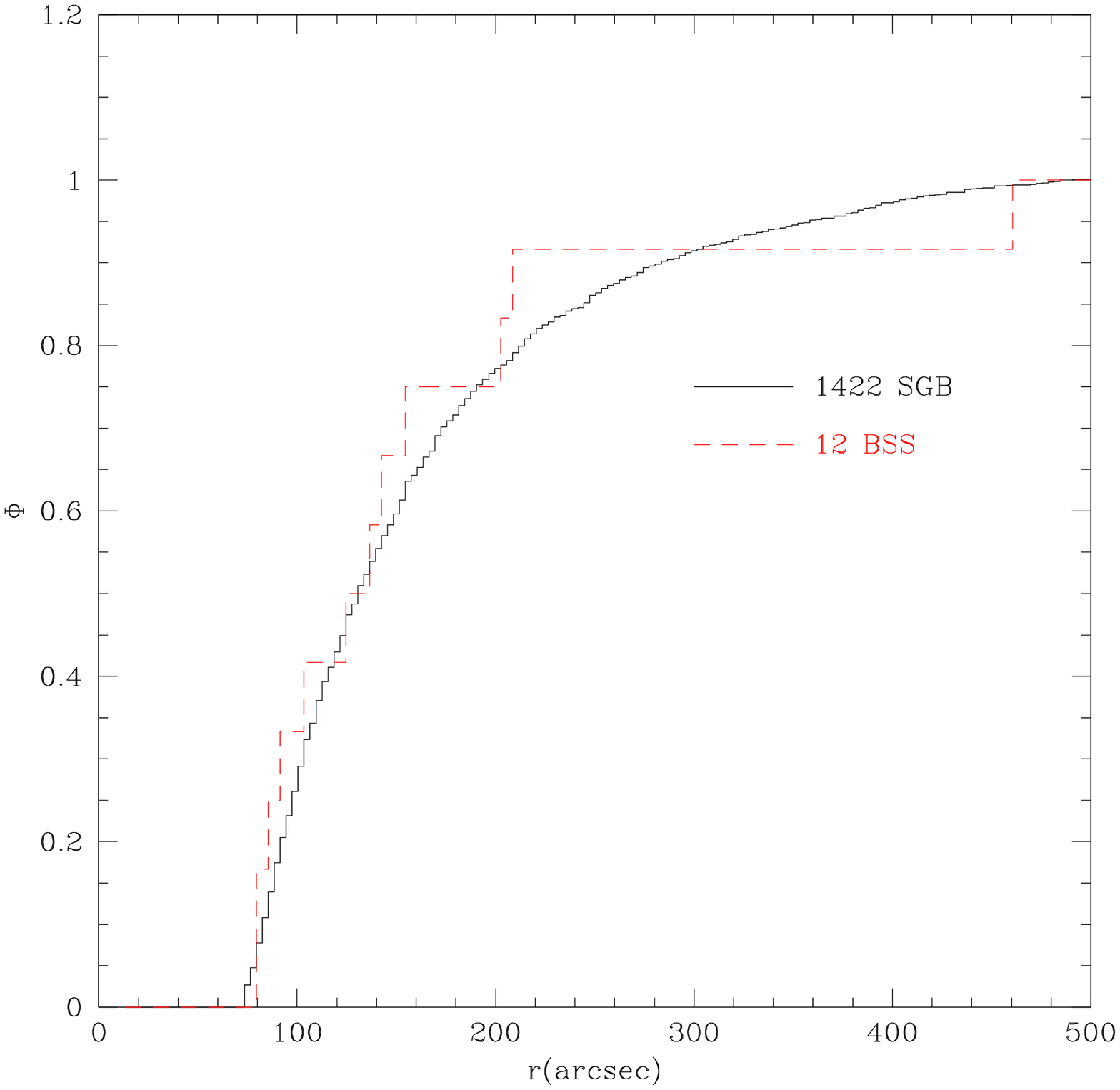]{Cumulative radial distributions for SGB stars
and BSS stars.
\label{fig6}}

\figcaption[Lee.fig07.ps]{Result of artificial star test for one direction
from cluster center. The difference $\Delta$ V is $V_{add} - V_{rec}$.
\label{fig7}}

\figcaption[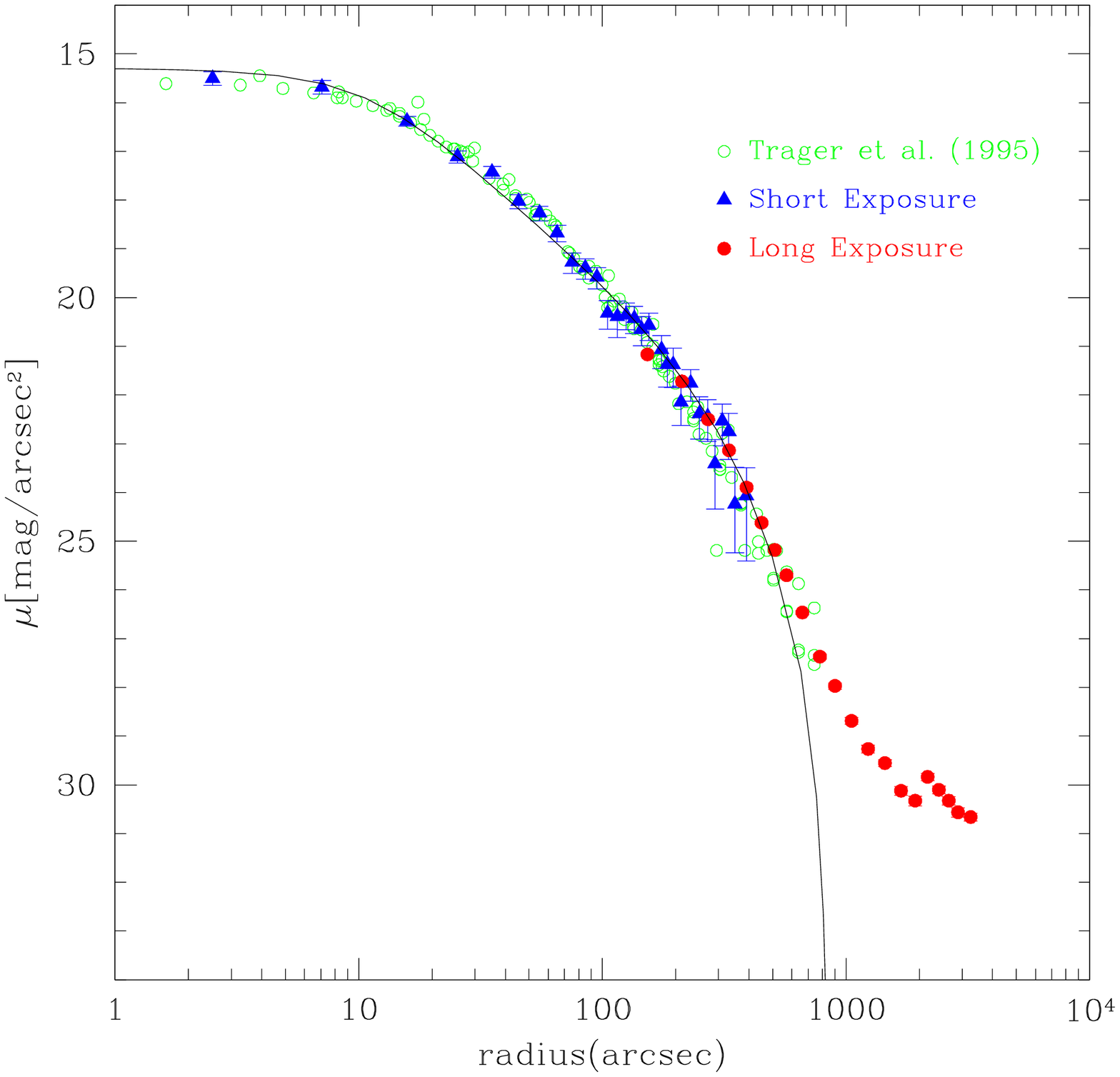]{Radial density profile of M92 using the entire
main sequence stars ($18.5 < V < 23.5$) within CMD criteria.
Filled triangles, short exposure data; filled circles, long exposure
data; open circles, Trager et al. (1995); solid line, single-mass
isotropic King model (W$_0$ = 8, c = 1.85).
\label{fig8}}

\figcaption[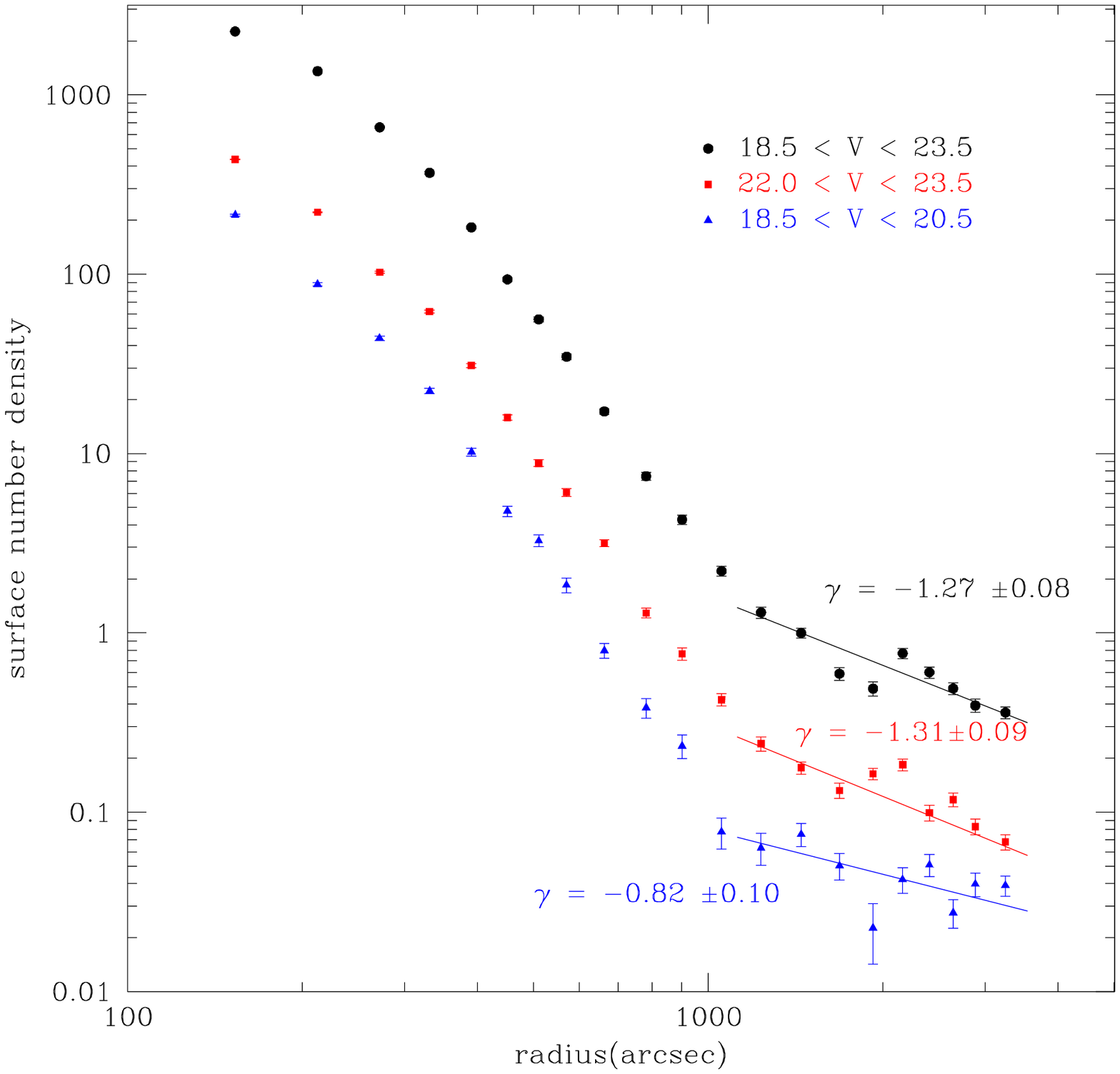]{Fit of power laws to the external profile of M92.
Triangles, the profile of bright stars ($18.5 < V < 20.5$); 
filled squares, the profile of faint stars ($22 < V < 23.5$); filled circles,
the profile of all stars ($18.5 < V < 23.5$); solid lines are the fitting
power laws with $\gamma = -0.82\pm0.10$, $\gamma = -1.31\pm0.09$, and $\gamma = -1.27\pm0.08$.
\label{fig9}}

\figcaption[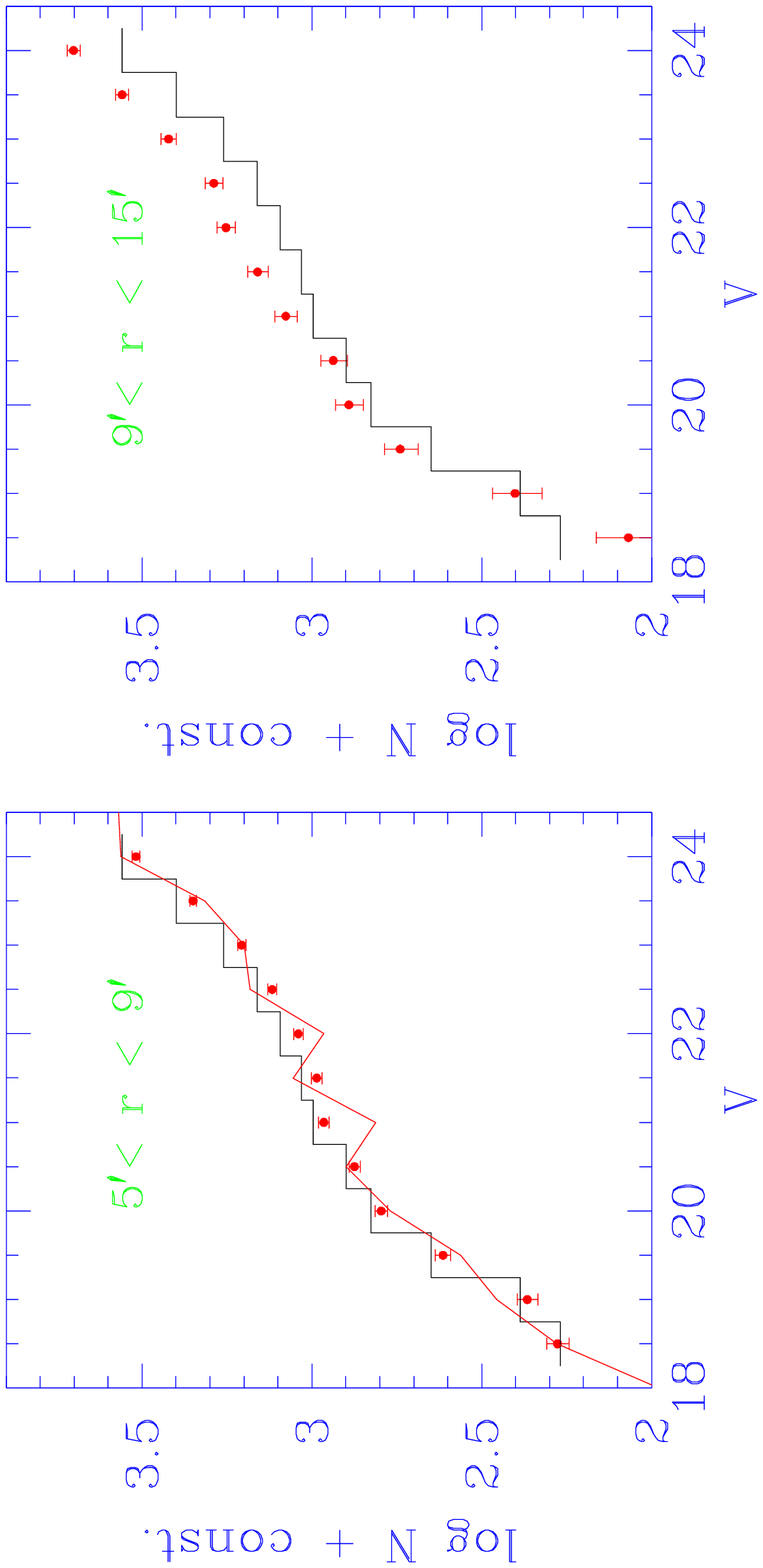]{Luminosity functions of inner ($5' < r < 9'$)
and outer ($9' < r < 15'$) region for M92.
The histogram is a LF of all stars within $5' < r < 15'$. 
A solid line overlayed with inner LF is the result from Piotto et al. (1997).
For convinient comparison, the LFs are arbitarily shifted.
\label{fig10}}

\figcaption[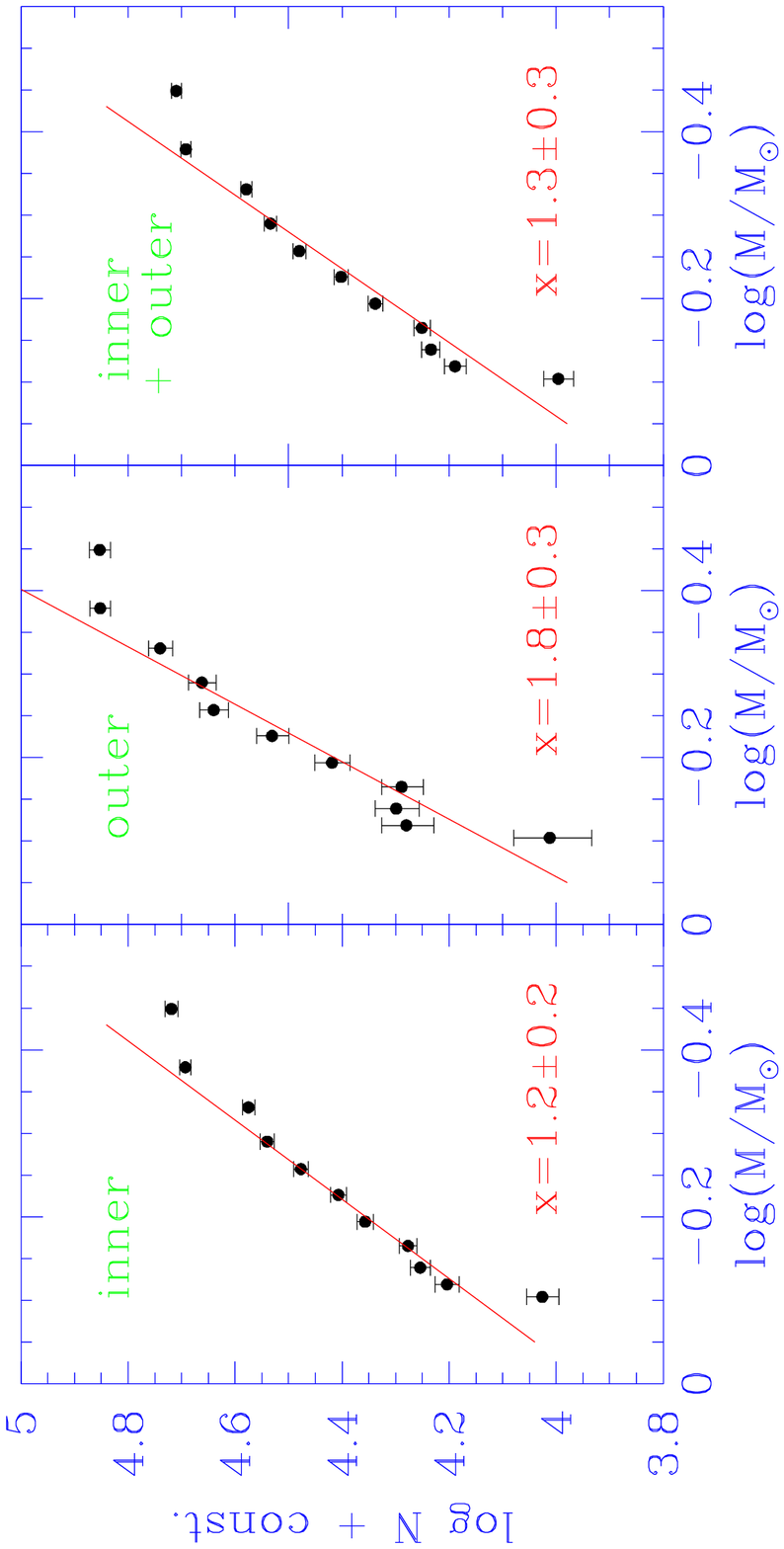]{Mass functions corresponding to the normalized luminosity
functions of Fig.~10. The lines show the indicated mass spectral index $x$ (see text).
\label{fig11}}

\figcaption[Lee.fig12.ps]{Surface density maps and contours of levels 
of different magnitude ranges ((a) $18.5 < V < 20$, (b) $18.5 < V < 23.5$).
The tidal radius is marked as a thick circle.
The two arrows indicate the directions of the galacic center (long one) and
compiled proper motion (Dinescu et al. 1999).
\label{fig12}}

\clearpage

\setcounter{table}{0}
\begin{deluxetable}{lrrrrrrrr}
\setlength{\tabcolsep}{3.5mm}
\tablecaption{Obsevational Information of M92 \label{tab1}}
\tablewidth{0pt}
\tablehead{
\colhead{Field} & \colhead{V-Filter} & \colhead{date} &
\colhead{I-Filter} & \colhead{date} }

\startdata
F1  & 2$\times$1200s & Jun. 28 2000 & 2$\times$900s & Jun. 27 2000 & \\
F2  & 2$\times$1200s & Jun. 28 2000 & 2$\times$900s & Jun. 27 2000 & \\
F3  & 3$\times$1200s & Jun. 28 2000 & 2$\times$900s & Jun. 28 2000 & \\
F4  & 3$\times$1200s & Jun. 28 2000 & 2$\times$900s & Jun. 28 2000 & \\
F5  & 3$\times$1200s & Jun. 28 2000 & 2$\times$900s & Jun. 29 2000 & \\
F6  & 3$\times$1200s & Jun. 28 2000 & 2$\times$900s & Jun. 29 2000 & \\
F7  & 3$\times$1200s & Jun. 29 2000 & 2$\times$900s & Jun. 29 2000 & \\
F8  & 2$\times$1200s & Jun. 29 2000 & 1$\times$900s & Jun. 29 2000 & \\
F9  & 2$\times$1200s & Jun. 29 2000 & 2$\times$900s & Jun. 29 2000 & \\
F10 & 2$\times$1200s & Jun. 29 2000 & 2$\times$900s & Jun. 29 2000 & \\
F1  & 2$\times$20s   & Jun. 27 2000 & 2$\times$20s  & Jun. 27 2000 & \\
F2  & 2$\times$20s   & Jun. 27 2000 & 2$\times$20s  & Jun. 27 2000 & \\
\enddata
\end{deluxetable}

\clearpage
\setcounter{table}{1}
\begin{deluxetable}{lrrrrrrrr}
\setlength{\tabcolsep}{3.5mm}
\tablecaption{Corrected Luminosity Functions and 
Incompleteness Correction Factors \label{tab2}}
\tablewidth{0pt}
\tablehead{
\colhead{V} & \multicolumn{2}{c}{Inner ($5'< r < 9'$)} & 
\multicolumn{2}{c}{Outer ($9' < r < 15'$)} &
\multicolumn{2}{c}{Inner + Outer} & \\
\colhead{} & \colhead{N} & \colhead{$f$} &
\colhead{N} & \colhead{$f$} & \colhead{N} & \colhead{$f$} }

\startdata
18.25-18.75 & 168.8 & 1.0 & 16.9 & 1.0 & 185.7 & 1.0 \\
18.75-19.25 & 207.0 & 1.0 & 36.4 & 1.0 & 243.4 & 1.0 \\
19.25-19.75 & 366.5 & 1.0 & 79.3 & 1.0 & 445.8 & 1.0 \\
19.75-20.25 & 557.5 & 1.0 & 112.4 & 1.0 & 669.9 & 1.0 \\
20.25-20.75 & 667.7 & 1.0 & 125.1 & 1.0 & 792.8 & 1.0 \\
20.75-21.25 & 822.0 & 0.97 & 172.6 & 1.0 & 989.7 & 0.98 \\
21.25-21.75 & 863.3 & 0.96 & 208.7 & 0.97 & 1074.2 & 0.96 \\
21.75-22.25 & 978.3 & 0.92 & 258.7 & 0.93 & 1239.8 & 0.92 \\
22.25-22.75 & 1167.2 & 0.92 & 278.2 & 0.92 & 1448.4 & 0.92 \\
22.75-23.25 & 1435.8 & 0.90 & 381.9 & 0.91 & 1821.9 & 0.90 \\
23.25-23.75 & 1999.5 & 0.79 & 523.1 & 0.86 & 2505.6 & 0.81 \\
23.75-24.25 & 3039.0 & 0.60 & 728.8 & 0.67 & 3556.5 & 0.65 \\
\enddata
\end{deluxetable}
\clearpage

\setcounter{table}{2}
\begin{deluxetable}{lrrrrrrrr}                                                          
\setlength{\tabcolsep}{2.0mm}
\tablecaption{Radial Density Profiles \label{tab3}}
\tablewidth{0pt}
\tablehead{
\multicolumn{4}{c}{Short Exposure ($15 < V < 20$)} 
& \multicolumn{4}{c}{Long Exposure($18.5 < V < 23.5$)} & \\
\colhead{log$r_{eff}$} & \colhead{\#/arcmin$^2$} & 
\colhead{SDP} & \colhead{error} &
\colhead{log$r_{eff}$} & \colhead{\#/arcmin$^2$} & 
\colhead{SDP} & \colhead{error} & \\
\colhead{(arcsec)} & \colhead{} & \colhead{(mag/arcsec$^2$)} & \colhead{} &
\colhead{(arcsec)} & \colhead{} & \colhead{(mag/arcsec$^2$)} & \colhead{} }

\startdata
 0.412 &   -   & 15.50 &  0.14 & 2.185 & 714.156 & 21.17 & 0.01 & \\
 0.849 &   -   & 15.68 &  0.14 & 2.327 & 428.225 & 21.72 & 0.01 & \\
 1.199 &   -   & 16.38 &  0.10 & 2.434 & 208.647 & 22.50 & 0.01 & \\ 
 1.406 &   -   & 17.11 &  0.12 & 2.520 & 116.156 & 23.14 & 0.02 & \\
 1.548 &   -   & 17.42 &  0.12 & 2.592 &  57.688 & 23.90 & 0.02 & \\
 1.656 &   -   & 18.02 &  0.15 & 2.654 &  29.572 & 24.62 & 0.03 & \\
 1.742 &   -   & 18.27 &  0.15 & 2.708 &  17.715 & 25.18 & 0.03 & \\
 1.814 & 0.268 & 18.68 &  0.17 & 2.756 &  10.972 & 25.70 & 0.04 & \\
 1.876 & 0.156 & 19.27 &  0.21 & 2.821 &   5.431 & 26.46 & 0.04 & \\
 1.930 & 0.139 & 19.39 &  0.21 & 2.893 &   2.358 & 27.37 & 0.05 & \\
 1.978 & 0.116 & 19.59 &  0.22 & 2.955 &   1.356 & 27.97 & 0.06 & \\
 2.022 & 0.059 & 20.32 &  0.30 & 3.023 &   0.699 & 28.69 & 0.07 & \\
 2.061 & 0.055 & 20.39 &  0.38 & 3.091 &   0.411 & 29.26 & 0.08 & \\
 2.097 & 0.058 & 20.35 &  0.28 & 3.160 &   0.316 & 29.55 & 0.07 & \\
 2.131 & 0.053 & 20.43 &  0.28 & 3.226 &   0.187 & 30.12 & 0.09 & \\
 2.162 & 0.044 & 20.65 &  0.30 & 3.284 &   0.155 & 30.33 & 0.10 & \\
 2.191 & 0.047 & 20.57 &  0.28 & 3.335 &   0.243 & 29.84 & 0.07 & \\
 2.243 & 0.030 & 21.07 &  0.34 & 3.381 &   0.190 & 30.10 & 0.08 & \\
 2.267 & 0.022 & 21.37 &  0.40 & 3.422 &   0.155 & 30.33 & 0.08 & \\
 2.290 & 0.022 & 21.37 &  0.40 & 3.460 &   0.125 & 30.56 & 0.09 & \\
 2.323 & 0.011 & 22.15 &  0.40 & 3.512 &   0.114 & 30.66 & 0.08 & \\
\enddata
\end{deluxetable}

\clearpage

\setcounter{table}{3}
\begin{deluxetable}{lrrrrrrrr}
\setlength{\tabcolsep}{2.0mm}
\tablecaption{Estimated Masses \label{tab4}}
\tablewidth{0pt}
\tablehead{
\colhead{$r(\arcsec)$} & \colhead{$V_{lim}$} &
\colhead{mesured ($M_\odot$)} &
\colhead{extrapolated ($M_\odot$)} & 
\colhead{total ($M_\odot$)} &
\colhead{slope ($x$)} }

\startdata
  0 - 14 & 22   & 3082 &  9216 & 12298 & 0.6 & \\
 14 - 28 & 22   & 5031 & 15041 & 20071 & 0.6 & \\
 28 - 42 & 22.5 & 4251 & 10051 & 14302 & 0.6 & \\
 42 - 56 & 23   & 5072 &  9294 & 14365 & 0.6 & \\
 56 - 70 & 23.5 & 5297 &  7432 & 12729 & 0.6 & \\
 70 - 84 & 24   & 5935 &  5647 & 11582 & 0.6 & \\
 84 - 98 & 24   & 4165 &  3957 &  8122 & 0.6 & \\
98 - 112 & 26   & 4021 &   891 &  4912 & 0.6 & \\
112 - 126 & 26  & 3946 &   868 &  4814 & 0.6 & \\
126 - 150 & 26.5 & 6459 & 1580 &  8039 & 1.0 & \\
150 - 200 & 26.5 & 9085 & 2180 & 11265 & 1.0 & \\
200 - 250 & 26.5 & 5975 & 1433 &  7409 & 1.0 & \\
250 - 300 & 26.5 & 4649 & 1116 &  5765 & 1.0 & \\
300 - 540 & 24   & 6028 & 11279 & 17308 & 1.2 & \\
540 - 900 & 24   & 1434 & 5498  &6931  &  1.8 & \\
\enddata 

\end{deluxetable}
\clearpage

\end{document}